\documentclass[aps,prd,a4paper,reprint,amsmath,amssymb,superscriptaddress,nofootinbib]{revtex4-2}

\usepackage{graphicx}
\usepackage{bm}
\usepackage[colorlinks=true, citecolor=teal]{hyperref}
\usepackage[capitalise]{cleveref}
\usepackage{booktabs}

\usepackage{anyfontsize}
\usepackage[utf8]{inputenc}
\usepackage{fourier}
\usepackage[scale=0.8]{GoMono}
\usepackage[T1]{fontenc}
\usepackage{microtype}

\newcommand{\rchi}{\raisebox{1.5pt}{\ensuremath{\chi}}}

\usepackage{listings}

\makeatletter
\lst@AddToHook{TextStyle}{\let\lst@basicstyle\ttfamily\normalsize}
\makeatother

\usepackage[dvipsnames,x11names]{xcolor}
\definecolor{white}{HTML}{FFFFFF}
\definecolor{black}{HTML}{000000}
\definecolor{blue}{HTML}{0000FF}
\definecolor{magenta}{HTML}{FF00FF}
\definecolor{violet}{HTML}{6A5ACD}
\definecolor{cyan}{HTML}{008A8C}
\definecolor{gray}{HTML}{BEBEBE}
\definecolor{green}{HTML}{2E8B57}
\definecolor{bordeaux}{HTML}{A52A2A}
\definecolor{red}{HTML}{FF0000}
\definecolor{yellow}{HTML}{FFFF00}
\definecolor{purple}{HTML}{A020F0}

\newenvironment{colornum}{
    \lstset{literate=%
        {0}{{{\color{magenta}0}}}1
        {1}{{{\color{magenta}1}}}1
        {2}{{{\color{magenta}2}}}1
        {3}{{{\color{magenta}3}}}1
        {4}{{{\color{magenta}4}}}1
        {5}{{{\color{magenta}5}}}1
        {6}{{{\color{magenta}6}}}1
        {7}{{{\color{magenta}7}}}1
        {8}{{{\color{magenta}8}}}1
        {9}{{{\color{magenta}9}}}1}
}
{
    \lstset{literate=%
        {0}{{{0}}}1
        {1}{{{1}}}1
        {2}{{{2}}}1
        {3}{{{3}}}1
        {4}{{{4}}}1
        {5}{{{5}}}1
        {6}{{{6}}}1
        {7}{{{7}}}1
        {8}{{{8}}}1
        {9}{{{9}}}1}
}

\makeatletter
\lstdefinestyle{cpp}{
    backgroundcolor=\color{white},
    commentstyle=\color{blue},
    keywordstyle=\color{green}\bfseries,
    numberstyle=\tiny\color{gray},
    stringstyle=\color{magenta},
    breakatwhitespace=false,
    breaklines=true,
    captionpos=b,
    keepspaces=true,
    numbers=left,
    numbersep=5pt,
    showspaces=false,
    showstringspaces=false,
    showtabs=false,
    tabsize=2,
    escapechar={|}, 
    basicstyle=%
    \ttfamily\normalshape
    \lst@ifdisplaystyle\footnotesize\fi
}
\makeatother

\newcommand{\code}[1]{\texorpdfstring{\lstinline{#1}}{#1}}

\usepackage{etoolbox}
\expandafter\patchcmd\csname \string\lstinline\endcsname{%
  \leavevmode
  \bgroup
}{%
  \leavevmode
  \ifmmode\hbox\fi
  \bgroup
}{}{%
  \typeout{Patching of \string\lstinline\space failed!}%
}

\newcommand{\neff}{n_\text{eff}}
\newcommand{\corneff}{\bar{n}_\text{eff}}
\newcommand{\like}{L}

\newcommand{\stdev}[1]{\sigma_{#1}}
\newcommand{\expected}[1]{\langle #1 \rangle}
\newcommand{\nMC}{n_\text{MC}}
\newcommand{\kMC}{k_\text{MC}}
\newcommand{\nLHC}{n_\text{LHC}}
\newcommand{\likeEstimator}{\hat\like}
\newcommand{\likeMLE}{\likeEstimator_\textsc{MLE}}
\newcommand{\likeUMVUE}{\likeEstimator_\text{UMVUE}}

\newcommand{\comb}[2]{{}^{#1}\!C_{#2}}
\newcommand{\given}{\,|\,}

\newcommand{\pmfPoisson}[2]{\text{Po}(#1 \given #2)}
\newcommand{\pmfBinom}[3]{\text{Binom}(#1 \given #2, #3)}

\newcommand{\Poisson}[1]{\text{Po}(#1)}
\newcommand{\Binom}[2]{\text{Binom}(#1, #2)}

\newcommand{\lumi}{\mathcal{L}}
\newcommand{\UMVUE}[1]{\text{UMVUE}\left[#1\right]}
\newcommand{\params}{\boldsymbol{x}}

\newcommand{\gev}{\,\text{GeV}}
\newcommand{\tev}{\,\text{TeV}}

\newcommand{\refcite}{ref.~\cite}

\newcommand{\firstrevise}[1]{#1}
\newcommand{\secondrevise}[1]{#1}
\newcommand{\thirdrevise}[1]{#1}
\newcommand{\fourthrevise}[1]{#1}

\begin{document}

\title{Bring the noise: exact inference from noisy simulations in collider physics}

\author{Christopher Chang}
\email{c.j.chang@fys.uio.no}
\affiliation{Department of Physics, University of Oslo, N-0316 Oslo, Norway}

\author{Benjamin Farmer}
\email{benf@gridcog.com}
\affiliation{Energy Modelling Group, Gridcog, VIC 3000, Australia}
\altaffiliation[Formerly ]{Department of Physics, Imperial College London,  London SW7 2AZ, United Kingdom}

\author{Andrew Fowlie}
\email{andrew.fowlie@xjtlu.edu.cn}
\affiliation{X-HEP Laboratory, Department of Physics, School of Mathematics and Physics, Xi'an Jiaotong-Liverpool University, Suzhou, 215123, China}

\author{Anders Kvellestad}
\email{anders.kvellestad@fys.uio.no}
\affiliation{Department of Physics, University of Oslo, N-0316 Oslo, Norway}

\begin{abstract}
We rely on Monte Carlo (MC) simulations to interpret searches for new physics at the Large Hadron Collider~(LHC) and elsewhere. These simulations result in noisy and approximate estimators of selection efficiencies and likelihoods. In this context we pioneer an exact-approximate computational method --- exact-approximate Markov Chain Monte Carlo (MCMC)\firstrevise{, also known as pseudo-marginal MCMC} --- that returns exact inferences despite noisy simulations. To do so, we introduce an unbiased estimator for a Poisson likelihood. We demonstrate the new estimator and new techniques in examples based on a search for neutralinos and charginos at the LHC using a simplified model. We find attractive performance characteristics --- exact inferences are obtained for a similar computational cost to approximate ones from existing methods and inferences are robust with respect to the number of events generated per point. \fourthrevise{The unbiased estimator uses a Poisson-distributed number of MC events; it is also possible to construct a biased estimator whose bias decays factorially with increasing number of MC events.}
\end{abstract}

\maketitle

\section{Introduction}

The ATLAS and CMS experiments at the Large Hadron Collider (LHC) are searching for new particles that are predicted by beyond the Standard Model (BSM) theories. The experimental signatures of the new particles typically depend on unknown parameters that control their masses and interactions~\cite{AbdusSalam:2020rdj}. There is, however, an underappreciated issue when using data from LHC searches to infer these unknown parameters. The data collected by the LHC and the LHC searches are so complicated that we cannot explicitly compute a likelihood function for use in a statistical analysis~\cite{Brehmer:2020cvb}.\footnote{Though see e.g., the matrix-element method~\cite{Volobouev:2011vb}.} Consequently, simplifications and approximations must be made.

Typically, we first make a lossy simplification of the experimental data: we discard the detailed event record and consider only whether an event falls into a judiciously-chosen signal region. The result is thus the number of events observed in the signal region. This is a counting experiment with a Poisson likelihood. However, this is not enough to allow us to compute the likelihood, as we cannot compute directly the expected number of such events from a model. Instead, we compute noisy estimators of the likelihood using Monte Carlo~(MC) simulations of the experiment on a computer, e.g., using the \code{Pythia}~\cite{Sjostrand:2014zea}, \code{Herwig}~\cite{Bahr:2008pv} and \code{Sherpa}~\cite{Sherpa:2019gpd} event-generators, and the \code{Delphes}~\cite{deFavereau:2013fsa} and \code{Geant}~\cite{Brun:1994aa} detector simulators. 

There are various statistical methodologies for learning about parameters using experimental data. We focus on Bayesian statistics, in which we learn about parameters by finding posterior distributions for them. The dominant algorithms for sampling from a posterior are variants of Markov Chain Monte Carlo (MCMC;~see e.g., \refcite{mcmc}). These algorithms are likelihood based; they require us to be able to compute the likelihood of the data observed by the LHC given a set of model parameters. Here, we pioneer exact-approximate MCMC --- an MCMC algorithm that returns exact results from noisy estimates of the likelihood.\footnote{Though note that other algorithms have exact-approximate variants, e.g., pseudo-marginal slice-sampling~\cite{2015arXiv151002958M}.}  \firstrevise{This class of algorithms is also known as pseudo-marginal MCMC.} For details of our numerical implementations in \code{Python}, \code{C++} and within the \code{ColliderBit}~\cite{GAMBIT:2017qxg} module of the \code{GAMBIT}~\cite{GAMBIT:2017yxo} framework, see \cref{sec:codes}. Our work differs from approximate simulation-based inference methods such as Approximate Bayesian Computation~(ABC; \cite{sisson2018}) in that it returns exact rather than approximate results.

\subsection{Exact-approximate MCMC}\label{sec:ea-mcmc}

Remarkably, in special circumstances approximate estimators of the likelihood can lead to exactly correct results; these algorithms are known as exact-approximate or, in the context of Bayesian computation, pseudo-marginal. In this work, we explore exact-approximate MCMC for LHC likelihoods. Exact-approximate MCMC was first applied in \refcite{Beaumont2003} and formalized in \refcite{Mller2006,Andrieu2009}\firstrevise{, where it is also referred to as \thirdrevise{pseudo}-marginal MCMC}. Similar ideas, however, were presented earlier in the context of lattice QCD~\cite{Lin:1999qu}. 

\firstrevise{Exact-approximate MCMC can be seen as MCMC that returns exact inferences despite a noisy estimator of the likelihood, $\likeEstimator$. In order for exact-approximate MCMC to work, the estimator must satisfy:}
\begin{enumerate}
    \item The estimator is unbiased (at least up to a constant factor) \firstrevise{with respect to the process by which it is generated during MCMC}, 
    \begin{equation}
        \expected{\likeEstimator} = C \like
    \end{equation}
    for constant $C$. That is, on average, the estimator equals the exact likelihood.
    \item The estimator is never negative,
    \begin{equation}
        \likeEstimator \ge 0.
    \end{equation}
    We later discuss adaptations of MCMC that remove this requirement.
\end{enumerate}
If the estimator satisfies these requirements, the MCMC algorithm converges to the exactly correct stationary distribution. When $\likeEstimator = \like$, this is standard MCMC. Unfortunately, the usual maximum likelihood estimator (MLE) for the LHC likelihood from MC simulations is biased.\footnote{The bias of the MLE for the selection efficiency was explored in~\refcite{2021arXiv211000294D}.} The cause of this bias is the difference in the expected distribution of events in the experiment versus the MC simulation. Running until a fixed number of events have accumulated results in binomially distributed MC samples, whereas the fixed-time experiment should expect Poisson-distributed samples. This means that MCMC would converge to the wrong stationary distribution. In this work, we develop a new unbiased estimator of the likelihood, and investigate its performance characteristics compared to the traditional estimator.

The fact that we use an estimator in exact-approximate MCMC leads to fluctuations in the likelihood. When an upwards fluctuations occurs, it becomes hard to accept subsequent transitions and the chain may become stuck. This leads to a poor MCMC acceptance rate and efficiency.\footnote{There are solutions such as clamping randomness~\cite{2015arXiv151002958M}.} Reducing the variance of the estimator, however, could be computationally expensive, and there is thus a trade-off between the variance of the estimator and stickiness in the chain. There are limited theoretical results about the optimal variance or performance of exact-approximate MCMC. In specific cases the optimal variance of $\log\likeEstimator$ was found to be around one~\cite{2012arXiv1210.1871D,Sherlock2015} and, as expected, the performance of exact-approximate MCMC converges to that of MCMC as the variance of the estimator goes to zero~\cite{2012arXiv1210.1484A}. Lastly, the optimal scaling of exact-approximate MCMC proposals differs from the well-known MCMC results~\cite{2014arXiv1408.4344S}, such that MCMC implementations may need to be re-tuned for optimal performance.

\section{Setting and estimators}

We consider counting experiments where we compare the expected and observed numbers of events in a particular region of phase space, called the signal region; see e.g., \refcite{GAMBIT:2017qxg} for an introduction. There is a known expected number of background events, $b$, and an unknown expected number of signal events, $s$.\footnote{We do not consider uncertainties on the background, $b$, in this work.} The experiment observes $o$ events with likelihood
\begin{equation}\label{eq:like}
    \like = \pmfPoisson{o}{\lambda} = \frac{e^{-\lambda} \lambda^o}{o!}
\end{equation}
where $\lambda = s + b$. The expected number of signal events, $s$, and thus the likelihood, are intractable functions of a model's parameters, $\params$. That is, we cannot evaluate $s(\params)$ directly. We may, however, express it as,
\begin{equation}
    s(\params) = \nLHC(\params) \times \epsilon(\params)
\end{equation}
where $\nLHC$ is the number of expected events from our model in the experiment and the selection efficiency, $\epsilon$, is the fraction of them that we expect to fall in the signal region. Fortunately, $\nLHC$ is tractable and may be computed through
\begin{equation}\label{eq:n_lhc}
    \nLHC(\params) = \lumi \times \sigma(\params)
\end{equation}
for an integrated luminosity $\lumi$ and production cross section $\sigma$. The selection efficiency, $\epsilon$, and thus the likelihood, may be \firstrevise{inferred} through MC. In MC we simulate signal-like events and their behavior in the experimental detector on a computer and count the fraction of events that fall into the signal region. \firstrevise{Inferences about the selection efficiency thus usually suffer from MC uncertainty.}

\firstrevise{There are several existing approaches to handling this MC uncertainty. These existing approaches treat the selection efficiency as a nuisance parameter and construct a joint likelihood describing the experiment and the MC simulations. In the classic Barlow-Beeston approach for frequentist inference~\cite{Barlow:1993dm}, the selection efficiency is subsequently profiled. This approach is further explored in \refcite{Arguelles:2019vrq,Dembinski:2022ios,Alexe:2024smm}.\footnote{\fourthrevise{We plan to explore advantages of our unbiased estimator of the likelihood in optimization and more generally in a frequentist context in future works.}} In the Bayesian approaches in \refcite{2013ICRC...33.3406C,Aggarwal:2011aa,Glusenkamp:2017rlp,Glusenkamp:2019uir,Arguelles:2019izp,Liu:2023bqc}, the selection efficiency is subsequently marginalized over a choice of prior.}

\firstrevise{Although these existing approaches incorporate MC uncertainty, the resulting inferences are inexact, as they reflect the results of finite-statistics MC simulations. In contrast, in exact-approximate MCMC, an estimator for the likelihood is implicitly averaged in a way that leads to exact inferences if the estimator is unbiased (referred to as pseudo-marginalization). In other words, if one uses an unbiased likelihood estimator, the inferences match what one would obtain if one could compute the efficiencies exactly. We now introduce two likelihood estimators: a common biased estimator and our new unbiased estimator.}

\subsection{Maximum likelihood estimator (MLE)}

Traditionally, \secondrevise{every time we wish to estimate a likelihood,} we generate a fixed number of MC events from our model, $\nMC$, with specialized software such as \code{Pythia}~\cite{Sjostrand:2014zea}. The computational cost of simulations is substantial and scales as $\mathcal{O}(\nMC)$. 
We count the number of simulated events that fall in the signal region, $k$. This follows a binomial distribution, $k \sim \Binom{\nMC}{\epsilon}$ with probability mass function (pmf)
\begin{equation}\label{eq:binom_pmf}
    \pmfBinom{k}{\nMC}{\epsilon} = \comb{\nMC}{k} \, \epsilon^k \, (1 - \epsilon)^{\nMC - k} 
\end{equation}
where we define $\comb{\secondrevise{n}}{k} = 0$ for $k > n$ or $k < 0$. From $k$ simulated events and \cref{eq:binom_pmf}, we construct the maximum likelihood estimator (MLE) for the unknown selection efficiency,
\begin{equation}
\hat\epsilon = \frac{k}{\nMC}. 
\end{equation}
We use this in \cref{eq:like} to construct an MLE estimate of the Poisson likelihood,
\begin{equation}
    \likeMLE = \pmfPoisson{o}{b + \hat \epsilon \nLHC} = \frac{e^{-(b + \hat \epsilon \nLHC)} (b + \hat \epsilon \nLHC)^o}{o!}.
\end{equation}
This MLE estimate is noisy and in fact biased, that is,
\begin{equation}
    \expected{\likeMLE} \equiv \sum_{k=0}^{\nMC} \likeMLE \, \pmfBinom{k}{\nMC}{\epsilon} \neq \like.
\end{equation}
It is, however, always positive and consistent, since $\likeMLE$ converges in probability to $\like$ as $\nMC \to \infty$. 

Typically, the number of events $\nMC$ is chosen so that the variance and bias are negligible. For $\epsilon \ll 1$, we may estimate the fractional uncertainty on the estimate of the selection efficiency by
\begin{equation}\label{eq:stdev_mle_efficiency}
    \frac{\stdev{\hat\epsilon}}{\hat\epsilon} =  \frac{\stdev{k}}{k} \simeq \frac{1}{\sqrt k},
\end{equation}
where $\stdev{}$ denotes standard deviation. Thus we might assume that simulating events until we collect $k \gtrsim 100$ signal events may be reasonable. The variance on the likelihood, however, depends on the ratio $\nLHC / \nMC$. E.g., consider the simple $o = 0$ case such that
\begin{equation}
    \log \likeMLE = -\frac{\nLHC}{\nMC} \, k - b
\end{equation}
with standard deviation,
\begin{equation}
    \stdev{\log \likeMLE} = \frac{\nLHC}{\nMC} \, \stdev{k}
    \propto \frac{1}{\sqrt {\nMC}}.
\end{equation}
Thus whilst we see the usual statistical scaling $1/\sqrt{\nMC}$, the ratio $\nLHC / \nMC$ plays an important role and we require $\nMC > \nLHC$ for negligible variance.

For LHC searches we are often fitting signals that are just a few events above background, that is, $s \simeq o - b \ll 100$. In this case, generating signal-like events until $k \gtrsim 100$ simultaneously implies that $\nMC > \nLHC$.

\subsection{Uniformly Minimum Variance Unbiased Estimator (UMVUE)}

\begin{figure}
    \centering
\begin{lstlisting}[language=Python,numbers=none,frame=single,framexrightmargin=-5pt,framexleftmargin=1.5pt,xleftmargin=5pt,framextopmargin=1pt]
def likelihood_estimate(n_mc, n_lhc, o, b, model_parameters):
    |\color{blue}\textrm{1.~Draw number of MC events from Poisson with mean $\nMC$}|
    k_mc = poisson_rvs(n_mc)
    |\color{blue}\textrm{2.~Generate $\kMC$ MC events using e.g.~\texttt{Pythia}}|
    events = generate_events(k_mc, model_parameters)
    |\color{blue}\textrm{3.~Count number of MC events that pass selections}|
    k = len(apply_selections(events))
    |\color{blue}\textrm{4.~Compute ratio $\nLHC / \nMC$}|
    f = n_lhc / n_mc
    |\color{blue}\textrm{5.~Return likelihood estimate using \cref{eq:UMVUE}}|    
    return umvue(k, f, o, b)
\end{lstlisting}
    \vspace{-3mm}
    \caption{\secondrevise{Pseudo-code for the UMVUE estimator of the likelihood. We write it here as a function of the user-chosen expected number of MC events, \code{n_mc}; the expected number of LHC events from \cref{eq:n_lhc}, \code{n_lhc}; the numbers of observed and background events in the search, \code{o} and \code{b}, respectively; and any model parameters for the signal model that must be used in MC event generation, \code{model_parameters}.}}
    \label{fig:pseudo_code_umvue}
\end{figure}

\firstrevise{For use in exact-approximate MCMC, we construct an estimator of the likelihood that is unbiased for any choice of $\nMC$.} In order to compute one, we slightly modify traditional MC simulation. \firstrevise{For complete mathematical details, see \cref{app:proof}.}
\secondrevise{For every likelihood estimate, we}
first draw the required number of MC events $\kMC$ from a Poisson distribution with mean $\nMC$, that is, $\kMC \sim \Poisson{\nMC}$. We then generate our MC events and denote the number that fall into the signal region by $k$, as before. This allows a simple unbiased estimator of the likelihood. When $b = 0$~\cite{Glasser1962},
\begin{equation}\label{eq:UMVUE_b0}
    \likeUMVUE = \pmfBinom{o}{k}{f} =
    \comb{k}{o} \, f^o \, (1 - f)^{k - o} 
\end{equation}
where $f = \nLHC / \nMC$. That is, the pmf for a binomial $o \sim \Binom{k}{f}$. This is not strictly a binomial, however, as the binomial success probability, here $f$, may be greater than one. In the more realistic $b > 0$ case,
\begin{equation}\label{eq:UMVUE}
\likeUMVUE = \sum\limits_{i=0}^{\secondrevise{o}} \pmfPoisson{o - i}{b} \,\pmfBinom{i}{k}{f}.
\end{equation}
When $k \sim \Poisson{\epsilon \nMC}$, these estimators are unbiased, $\expected{\likeUMVUE} = \like$, as shown in \cref{app:proof} by a power-expansion similar to that in \refcite{Glasser1962}. \secondrevise{For clarity, we show pseudo-code for producing this estimator in \cref{fig:pseudo_code_umvue}.}

When $f=1$, \cref{eq:UMVUE_b0} reduces to an indicator function that requires the simulated number of events to exactly equal the observed number of events. This resembles ABC with zero tolerance. The fact that \cref{eq:UMVUE} is a binomial probability was remarked upon in \refcite{Glasser1962}. Significantly, it means that in the limiting case $\nMC \to \infty$, the UMVUE estimator approaches the MLE.

The benefits of using $\kMC \sim \Poisson{\nMC}$ rather than a fixed number of MC events are twofold. First, as shown in \cref{app:no_unbiased_binomial}, an unbiased estimate does not exist if we simulate a fixed number of MC events. Second, it trivially generalizes to cases where the likelihood is a product of Poisson likelihoods,
\begin{equation}\label{eq:product}
    \left\langle \prod\nolimits_i \likeUMVUE {}_i \right\rangle = 
    \prod\nolimits\nolimits_i \left\langle \likeUMVUE {}_i \right\rangle =
    \prod\nolimits_i \like_i,
\end{equation}
even if the same MC simulations are used to estimate each factor. This trivial generalization happens because the number of simulated events in each search follow independent $\Poisson{\epsilon_i \nMC}$ distributions. \fourthrevise{As discussed in \cref{app:no_unbiased_binomial}, however, using a fixed number of MC events, $\nMC$, it would still be possible to construct an UMVUE of the first $\nMC$ terms in an expansion of the Poisson likelihood. The residual bias would decay factorially with increasing $\nMC$.}

\secondrevise{As shown in \cref{app:proof}, the estimators \cref{eq:UMVUE_b0,eq:UMVUE} are the uniformly minimum variance unbiased estimators (UMVUE) of the likelihood possible from this procedure. Despite that, they suffer from undesirable properties. E.g., the estimator \cref{eq:UMVUE_b0} can be zero when $k < o$ and even negative when $f > 1$. In these cases the vanishing bias of \thirdrevise{$\likeUMVUE$ originates} from cancellations between \thirdrevise{overestimates, $\likeUMVUE \gg L$, and underestimates, $\likeUMVUE \ll L$,  of the likelihood}. Thus, even though it is the UMVUE, the variance can be enormous. Whilst variance cannot spoil the correctness of exact-approximate algorithms, as discussed in \cref{sec:ea-mcmc}, it could make them inefficient and it cannot be taken for granted that these estimators will work well in practice. Thus, we now explore the performance characteristics of these estimators in numerical experiments.} 

\thirdrevise{The existence of these ``absurd'' UMVUE estimators is discussed in \refcite{lehmann1983} and attributed to the fact that we are ``estimating the probability of an event
of interest occurring over a long period of time from observations over a much shorter period.'' That is indeed the case here, with the number of events playing the role of time: when $f > 1$, we are estimating the probability of $o$ events in $\nMC$ events from observations of only $\nLHC < \nMC$ events. There are relevant examples in \refcite[Example 3.11]{lehman1986} and \refcite[Example 17.26]{kendall1979}.}

\section{Methods}

\subsection{Negative estimates of likelihood}

When $f > 1$, the UMVUE may be negative. This issue is sometimes known as the ``sign problem''~\cite{Lin:1999qu}. To use this estimator in an exact-approximate MCMC framework, we follow \refcite{Lyne2015} and take the absolute value of the estimator, $|\likeUMVUE|$, but save the sign, 
\begin{equation}
    \sigma = \begin{cases}
        +1 & \likeUMVUE \ge 0\\
        \thirdrevise{-1} & \likeUMVUE < 0
    \end{cases}
\end{equation}
to the chain. When computing expectations, estimating densities, or computing effective sample size (ESS), we take into account the sign using weighted sums. 

Indeed, the corrected ESS for parameter $x$, denoted $\corneff(x)$, depends on the sign through
\begin{equation}\label{eq:neff_sigma}
    \corneff(x) = \left[\frac{\sum_{i=1}^n \sigma_i}{n}\right]^2 \, \neff(\sigma x),
\end{equation}
where $n$ is the number of samples in the chain and $\neff(\sigma x)$ is the uncorrected ESS for the parameter $\sigma x$. That is, the ESS for $x$ is the ESS for the weighted quantity $\sigma x$ reduced by a factor accounting for the negative weights. When there are as many negative as positive sign likelihood samples, there are effectively no samples and $\corneff = 0$. 

\subsection{Performance metrics}

In the past even when the maximum likelihood estimator was used in MCMC, it was not knowingly done so in an exact-approximate MCMC framework. Instead, substantial $\nMC$ was chosen to make variance and bias negligible and justify an approximation $\hat L \approx L$. We know of no general analytic results about the relative performance of these two estimators in exact-approximate MCMC (though see \refcite{2014arXiv1404.6909A}). Broadly speaking, we anticipate that the UMVUE may perform better, as we do not need to increase $\nMC$ to remove the bias in our estimator; instead, the bias averages out over the whole chain. On other hand, the UMVUE estimator is far noisier, and this degrades the efficiency of exact-approximate MCMC as it results in sticking behavior and increased auto-correlation.

We judge performance by the effective number of posterior samples obtained --- effective sample size (ESS) --- per MC event simulated,
\begin{equation}
    \text{ESS per MC event} = \frac{\text{ESS}}{\text{Total number of MC events}}.
\end{equation}
We run an affine-invariant MCMC ensemble sampler~\cite{Goodman2010}, \code{emcee}~\cite{Foreman-Mackey:2012any}, with 10 walkers for \thirdrevise{at least} 100\,000 steps. 
The bulk ESS is computed using \code{ArviZ}~\cite{arviz_2019}, adjusted via \cref{eq:neff_sigma} when necessary. 

\subsection{Toy problems}

We test the methods on toy problems. We take the luminosity, numbers of observed and background events from the \code{SRWZ_15} signal region of the \code{ATLAS-SUSY-2019-09} search~\cite{ATLAS:2021moa}.
That is, $o = 5$, $b = 2.8$ and $\lumi = 139/\text{fb}$ of data. This $\sqrt{s} = 13\tev$ search targeted neutralino-chargino production in final states with three leptons and missing transverse momentum. 

We consider three toy models with increasing realism:
\begin{enumerate}
    \item First, we construct a simple one-dimensional model in which the selection efficiency, $\epsilon$, is an input parameter. We take a flat prior on $\epsilon$ and fix the cross section to $\sigma = 1000\,\text{fb}$. 
    
    \item Second, we construct a simplified model based on the  \code{TChiWZ} topology: mass-degenerate  $\rchi^\pm_1$ and $\rchi_2$ particles are pair produced and decay exclusively through $\rchi^\pm_1 \to W \rchi_1$ and $\rchi_2 \to Z \rchi_1$, respectively. 

    We fix the production cross section to $\sigma = 1000\,\text{fb}$ and compute a selection efficiency for the \code{SRWZ_15} signal region as a function of $(m_1, m_2)$ through \code{SModelS}~\cite{Alguero:2021dig}.
    
    There are thus two unknown parameters: $m_1 \equiv m(\rchi_1)$ and $m_2 \equiv m(\rchi_2) = m(\rchi^\pm_1)$. We take flat priors between $0 < m_1 < 300\gev$ and $m_1 + M_Z < m_2 < 300\gev$.
    
    \item Lastly, we use the same simplified model, but compute the production cross-section $\sigma(pp\to\rchi^{\phantom\pm}_2\rchi^\pm_1)$ as a function of $(m_1, m_2)$ using tabulated wino-like neutralino-chargino production cross-sections~\cite{xs,Fuks:2012qx,Fuks:2013vua}. 
    
    The tabulated cross sections were computed at NLO + NLL precision assuming mass-degenerate wino-like $\rchi_2$ and $\rchi^{\pm}_1$, and a bino-like $\rchi_1$ and use an envelope of \code{CTEQ6.6} and \code{MSTW2008nlo90cl} parton distribution functions~\cite{xs}.
\end{enumerate}

\secondrevise{In our toy problems,} we never simulate collider events with e.g., \code{Pythia} as this is computationally expensive and unnecessary for our purposes. Instead, we perform white-box experiments in which we internally choose a true selection efficiency $\epsilon$ and use it as a parameter in a binomial or Poisson distribution for the MLE and UMVUE estimators, respectively. \secondrevise{At every point}, we then simulate from the binomial and Poisson distributions directly, rather than perform collider simulations. 

\section{Results}

\begin{figure}[t]
    \centering
    \includegraphics[width=0.9\linewidth]{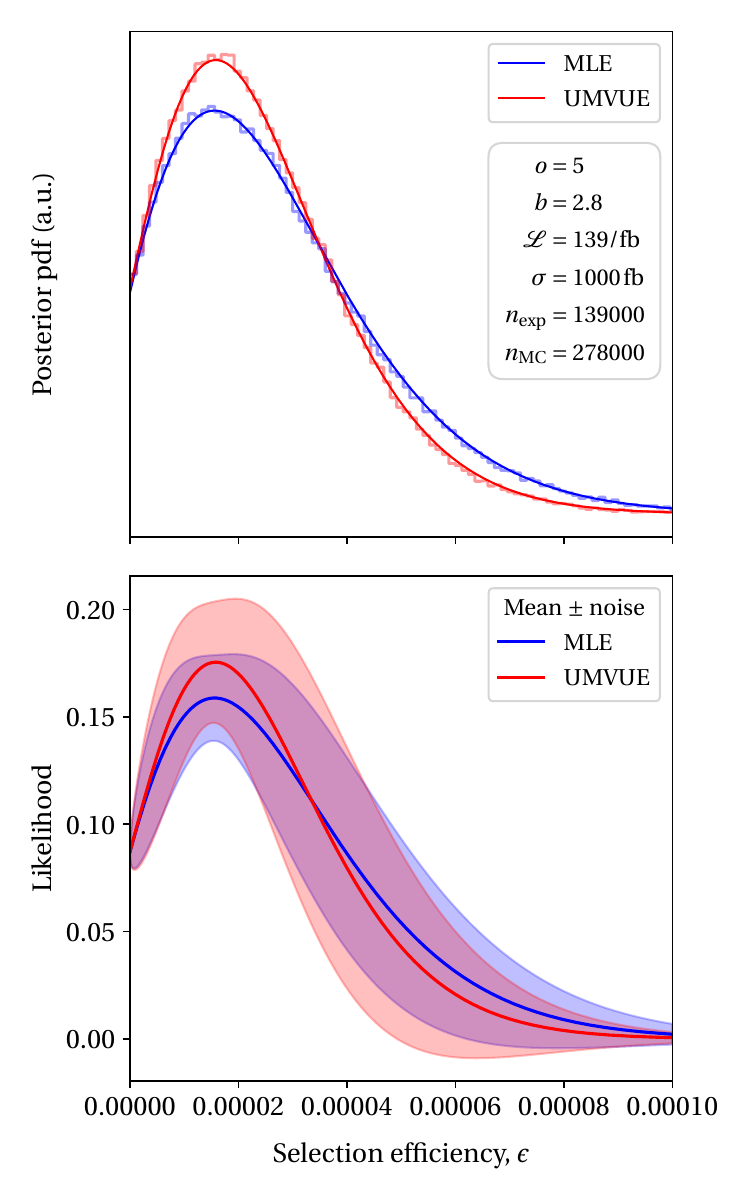}
    \caption{The posterior pdf reconstructed by MCMC for the unknown selection efficiency (top) and the mean and standard deviation of the likelihood function (bottom) using MLE and UMVUE Poisson likelihood estimators.}
    \label{fig:eff_1d}
\end{figure}

In \cref{fig:eff_1d} we show the posterior pdf (upper panel) and likelihood estimates (lower panel) for the unknown selection efficiency, $\epsilon$, for $\nMC = 2 \nLHC$. The MLE and UMVUE estimators produce samples at similar rates, around one posterior sample per $10^7$ MC events. The MCMC results (stepped histogram) match the expected results (smooth lines). The MLE estimator, however, shows noticeable bias in the posterior. The mean and standard deviation of the likelihood estimates (lower panel) show that the estimators are similarly noisy; though for the UMVUE, at equilibrium that noise results in the exactly correct distribution.

\begin{figure}[t]
    \centering
    \includegraphics[width=0.9\linewidth]{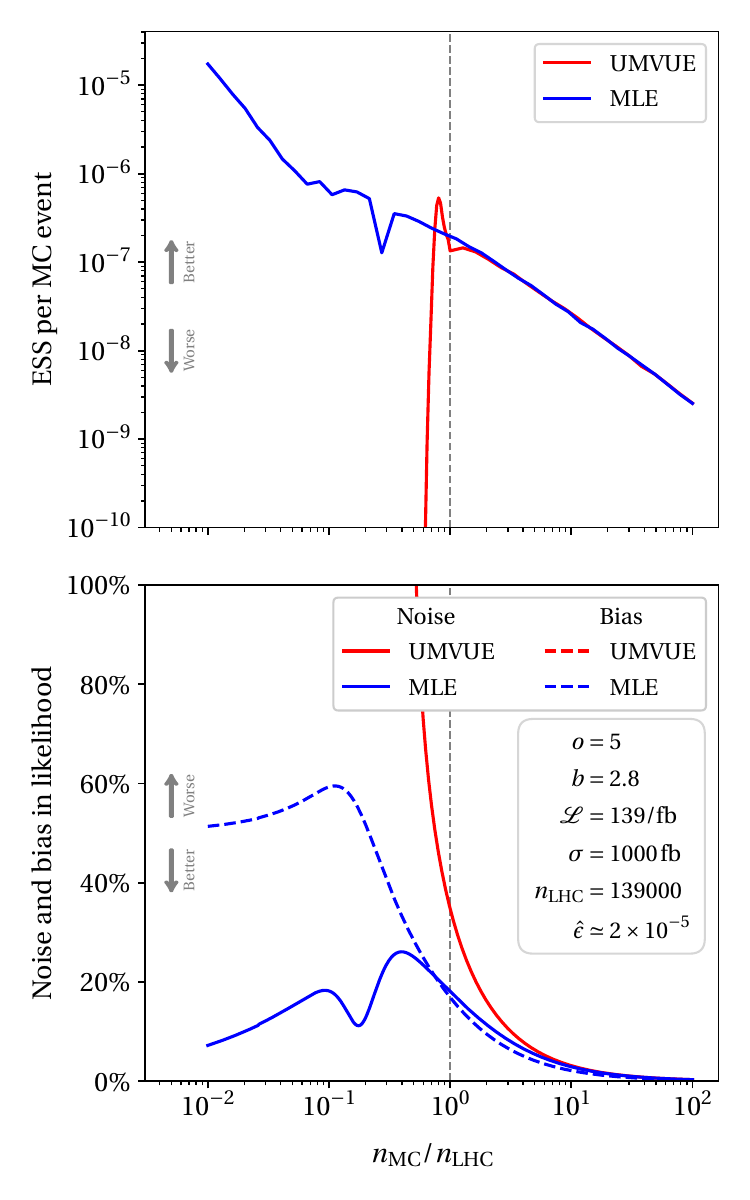}
    \caption{Effective number of posterior samples for the unknown selection efficiency per MC simulation (top) and bias and variance (bottom) for the MLE and UMVUE Poisson likelihood estimators relative to the exact likelihood as we change $\nMC$.}
    \label{fig:var_f}
\end{figure}

In \cref{fig:var_f} we explore the performance on this problem as we change the number of MC events. First, consider the MCMC efficiency (top panel). For $\nMC < \nLHC$, the UMVUE estimator suffers from terrible efficiency, due to the possibility of negative likelihood estimates. The MCMC efficiency itself becomes noisy and hard to estimate reliably. Once $\nMC \gtrsim \nLHC$, however, the noise becomes unimportant and the UMVUE and MLE estimators show remarkably similar MCMC efficiencies that deteriorate as $1 / \nMC$. \firstrevise{Thus, the UMVUE estimator yields unbiased inferences for approximately the same computational cost as biased ones.} For the UMVUE, the best choice appears to be about $\nMC \simeq \nLHC$. For the MLE estimate, $\nMC$ should be as small as possible whilst resulting in an acceptable bias. 

We show the noise and bias (lower panel) at the best-fit selection efficiency. For the MLE estimator, the noise goes to zero as number of simulations goes to zero, as we always estimate that $k = 0$. Similarly bias reaches a horizontal asymptote: the bias of the $k = 0$ estimate. As the number of simulations increases, the noise and bias decay as $1/\sqrt{\nMC}$. For the UMVUE estimator, the bias always vanishes. As discussed, the noise starts as enormous when $\nMC < \nLHC$, and subsequently decays as $1/\sqrt{\nMC}$.

\begin{figure}[t]
    \centering
    \includegraphics[width=0.9\linewidth]{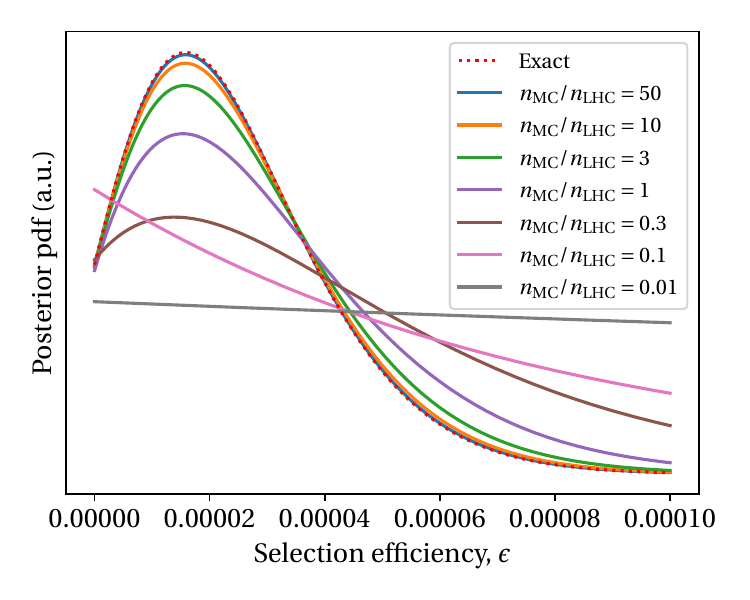}
    \caption{The posterior pdf for the selection efficiency from the MLE likelihood estimator as we change $\nMC$. We show the exact result (red) for reference.}
    \label{fig:post_var_f}
\end{figure}

Lastly, we consider the impact of the bias on inferences in \cref{fig:post_var_f} by showing the posterior pdf for the selection efficiency from the MLE likelihood estimator as we change $\nMC$. We see that for $\nMC < \nLHC$, the pdf from the MLE estimator is much broader and flatter than the exact one. When $\nMC / \nLHC = 0.01$, the pdf is almost flat, resembling the prior for the selection efficiency. Thus, the bias when $\nMC < \nLHC$ leads to faulty inferences. Indeed, for a posterior that is indiscernible by eye from the true one in \cref{fig:post_var_f}, requires $\nMC / \nLHC \gtrsim 50$. This choice would perform a factor of about $25$ worse than the UMVUE estimator at $\nMC / \nLHC = 2$.

\begin{figure}[t]
    \centering
    \includegraphics[width=0.99\linewidth]{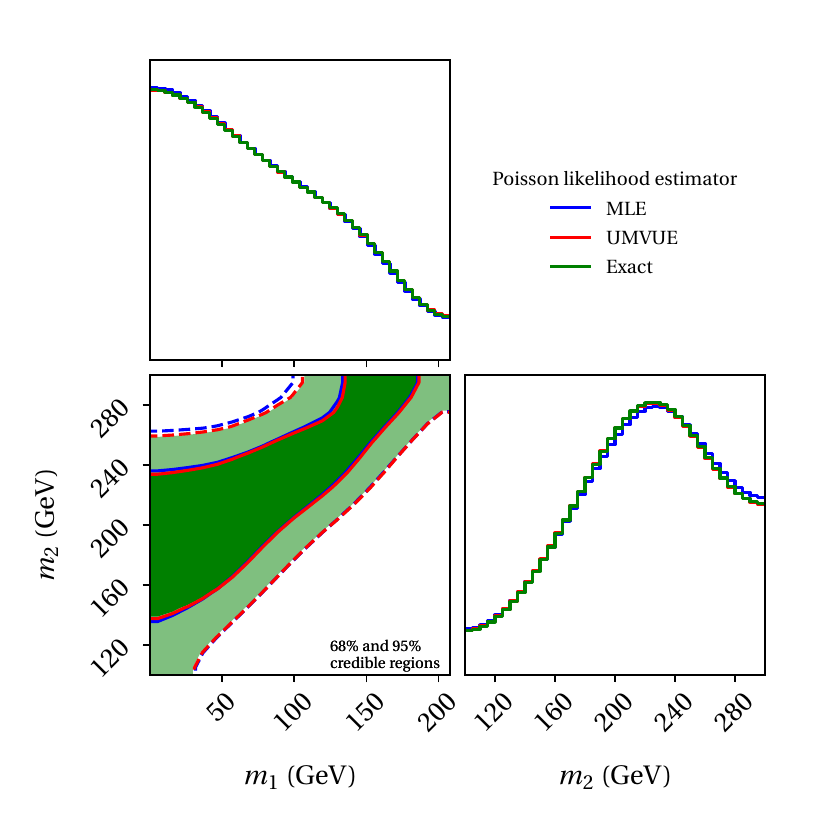}
    \caption{Posterior pdf for the \code{TChiWZ} simplified model masses and \code{SRWZ_15} signal region of the \code{ATLAS-SUSY-2019-09} search~\cite{ATLAS:2021moa} found from MLE and UMVUE Poisson likelihood estimators, and the exact Poisson likelihood.}
    \label{fig:m1m2_corner_1}
\end{figure}

We now consider the \code{TChiWZ} model in which degenerate $\rchi_2\rchi^\pm_1$ particles are pair produced and decay through $\rchi_2 \to Z \rchi_1$ and $\rchi^\pm_1 \to W \rchi_1$, respectively. There are two unknown parameters, $m_1$ and $m_2$, that determine the selection efficiency. We run 10 walkers for \thirdrevise{1\,000\,000} steps \thirdrevise{and $\nMC = 2 \nLHC$}. \thirdrevise{In each case, we collect an ESS of about 200\,000 samples and obtain $\hat R = 1.0$~\cite{vehtari2021}}. In \cref{fig:m1m2_corner_1} we show $68\%$ and $95\%$ credible regions on the $(m_1, m_2)$ plane, as well as one-dimensional marginal posterior distributions, computed from the exact likelihood and from our two estimators. \thirdrevise{The MLE estimator and UMVUE estimator result in about $10^{-7}$ samples per MC event for both the $m_1$ and $m_2$ parameters.}
The UMVUE estimator agrees with the exact result, as expected, though the MLE estimator shows bias. The differences between the MLE and exact result are slight though noticeable, particularly in the 95\% contours on the $(m_1, m_2)$ plane. \thirdrevise{We present estimates of the posterior means and MC standard error (MCSE)  in \cref{tab:m1m2_1}. As expected the UMVUE estimator agrees with results from the exact likelihood, whereas the MLE shows significant bias. Lastly, we perform Kolmogorov-Smirnov (KS) two-sample tests~\cite{kolmogorov1933,smirnov1948} on the $m_2$ samples thinned to the ESS. The MLE and UMVUE samples are significantly different ($p < 10^{-9}$) whereas the UMVUE and exact likelihood samples have no significant difference ($p \simeq 0.1$).} 

\begin{table}[t]
    \centering
    \begin{tabular}{lcc}
        \toprule
        Estimator &  \multicolumn{2}{c}{Posterior mean $\pm$ MCSE} \\
        \midrule
        & $m_1$ (GeV) & $m_2$ (GeV)\\
        \cmidrule(lr){2-3}
        MLE & $76.0 \pm 0.1$ & $219.4 \pm 0.1$\\
        UMVUE & $76.4 \pm 0.1$ & $218.8 \pm 0.1$\\
        Exact & $76.4 \pm 0.1$ & $219.0 \pm 0.1$\\
        \bottomrule
    \end{tabular}
    \caption{\thirdrevise{Mean and MC standard error for the \code{TChiWZ} simplified model masses and \code{SRWZ_15} signal region of the \code{ATLAS-SUSY-2019-09} search~\cite{ATLAS:2021moa} found from MLE and UMVUE Poisson likelihood estimators, and the exact Poisson likelihood.}}
    \label{tab:m1m2_1}
\end{table}

\begin{figure}[t]
    \centering
    \includegraphics[width=0.99\linewidth]{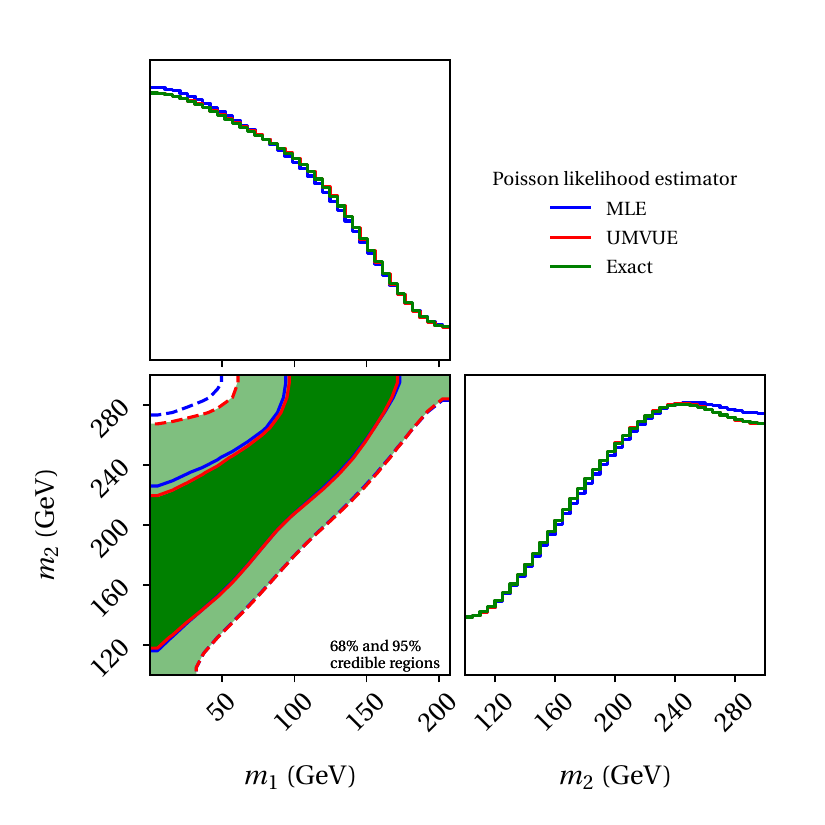}
    \caption{As in \cref{fig:m1m2_corner_1}, though with the cross-section computed as a function of $(m_1, m_2)$ at NLO + NLL.}
    \label{fig:m1m2_corner}
\end{figure}

Finally, we compute the cross section as a function of $(m_1, m_2)$. As before, we run 10 walkers for \thirdrevise{1\,000\,000} steps \thirdrevise{and $\nMC = 2 \nLHC$, obtaining an ESS of about 200\,000 and $\hat R = 1.0$~\cite{vehtari2021}}. \thirdrevise{The} MLE estimator and UMVUE estimator result in about $10^{-7}$ \thirdrevise{samples per MC event} for both the $m_1$ and $m_2$ parameters. We show $68\%$ and $95\%$ credible regions on the $(m_1, m_2)$ plane, as well as one-dimensional marginal posterior distributions in \cref{fig:m1m2_corner}. \thirdrevise{The shapes of the distributions slightly changed, although remain similar to \cref{fig:m1m2_corner_1}. The MLE estimator shows slightly more bias in the $(m_1, m_2)$ parameters in \cref{fig:m1m2_corner_1}.} \thirdrevise{The bias can be compared to the MCSE in \cref{tab:m1m2}. Lastly, we again perform a KS test on the $m_2$ samples. The MLE and UMVUE samples are significantly different ($p < 10^{-9}$) whereas the UMVUE and exact likelihood samples have no significant difference ($p \simeq 0.5$).} 

\begin{table}[t]
    \centering
    \begin{tabular}{lccc}
        \toprule
        Estimator &  \multicolumn{3}{c}{Posterior mean $\pm$ MCSE} \\
        \midrule
        & $m_1$ (GeV) & $m_2$ (GeV) & $\sigma$ (fb)\\
        \cmidrule(lr){2-4}
        MLE   & $74.3 \pm 0.1$ & $223.2 \pm 0.1$ & $2082 \pm 5$\\
        UMVUE & $75.1 \pm 0.1$ & $222.2 \pm 0.1$ & $2099 \pm 5$\\
        Exact & $75.2 \pm 0.1$ & $222.2 \pm 0.1$ & $2103 \pm 5$\\
        \bottomrule
    \end{tabular}
    \caption{\thirdrevise{As in \cref{tab:m1m2_1}, though with the cross-section computed as a function of $(m_1, m_2)$ at NLO + NLL, and shown as well as the mass parameters.}}
    \label{tab:m1m2}
\end{table}

\section{Discussion and conclusions}

The Poisson likelihood in collider searches for new physics can only be estimated through noisy MC simulations. Our inferences are thus at present noisy and approximate, though the noise can be reduced by increasing the number of MC events. With this in mind, we pioneered exact-approximate inference in this setting --- exact statistical inference despite an approximate likelihood. The exact-approximate MCMC algorithm \firstrevise{(also known as \thirdrevise{pseudo}-marginal MCMC)} required us to a construct a novel, unbiased estimator for a Poisson likelihood. The estimator is only slightly more complicated than the  traditional maximum likelihood one, and we provide public \code{C++}, \code{Python} and \code{ColliderBit} implementations.

We investigated the MCMC efficiency, bias and noise characteristics of the traditional and new estimators on toy problems. The toy problems were based on an ATLAS search for new physics with the \code{TChiWZ} topology. We constructed a simplified model with unknown masses of a parent and daughter particle. Lastly, we assumed that these were chargino and neutralino particles and computed realistic cross sections using tabulated NLO + NLL predictions. We found that, with our new unbiased estimator, exact results could be obtained at a remarkably similar computational cost to biased results. The performance, however, depended strongly on the average number of MC events that were simulated per point. The unbiased estimator suffered from terrible noise and inefficiency when an inadequate number of MC events was chosen. Thus, the most efficient choice of estimator depends on the degree of bias that a user can tolerate. Lastly, the unbiased estimator is undoubtedly a safer choice as it returns exact inferences regardless the number of MC simulations chosen; whereas the MLE estimator returns faulty inferences when the number of MC simulations was inappropriately chosen.

There remains a few open questions. There may be other estimators with advantageous bias-variance characteristics. We anticipate that a degree of bias could be tolerated if the variance was such that it led to improved performance. In particular, it might be possible, though challenging, to construct a non-negative unbiased estimator~\cite{10.1214/15-AOS1311}, or there may be ways to reduce the bias in the MLE estimator to a tolerable level~\cite{2013arXiv1311.6311K}. 
\fourthrevise{Lastly, if one wished to stick to a fixed number of MC events, it would be possible to construct a biased estimator whose bias decays factorially with an increasing number of MC events.}
For now, though, the estimator and methods here show acceptable performance and result in exact inferences, and could become standard in Bayesian inference involving collider likelihoods or more generally noisy estimates of Poisson likelihoods. 

\begin{acknowledgments}
We would like to thank Andy Buckley for discussions and feedback. 
AF was supported by RDF-22-02-079 and NSFC RFIS-II W2432006. CC and AK were supported by the Research Council of Norway (RCN) through the FRIPRO grant 323985 PLUMBIN’.
\end{acknowledgments}

\clearpage

\bibliographystyle{JHEP}
\bibliography{refs}

\clearpage
\appendix
\onecolumngrid

\section{Proof that estimator is UMVUE}\label{app:proof}

We wish to show that our estimator is the uniform minimum variance unbiased estimator (UMVUE). That is, it is expected to equal the true likelihood when $k \sim \Poisson{\epsilon \nMC}$, and that the variance of our estimator is the minimum among estimators with this property for all possible values of the likelihood. The expectation of our estimator may be written,
\begin{equation}
    \langle \likeUMVUE \rangle = \sum\limits_{k=0}^\infty \likeUMVUE \, \pmfPoisson{k}{\epsilon \nMC},
\end{equation}
and the exact likelihood is a Poisson,
\begin{equation}
      \like = \pmfPoisson{o}{b + \epsilon \nLHC}.
\end{equation}
First consider the $b = 0$ case; demonstrating that \cref{eq:UMVUE_b0} is the UMVUE is extremely similar to the proof in~\refcite{Glasser1962}. First, we write the Poisson likelihood as a series in $s$,
\begin{equation}\label{eq:series}
 \pmfPoisson{o}{s}  = \frac{e^{-s} s^o}{o!} =
 \sum_{i=0}^\infty \frac{(-1)^i}{i!o!} \, s^{o +i}.
\end{equation}
We now construct the UMVUE by replacing each power of $s$ by the UMVUE,
\begin{equation}\label{eq:subst}
    \UMVUE{\pmfPoisson{o}{s}} = \left.\pmfPoisson{o}{s}\right|_{s^n \to \UMVUE{s^n} }.
\end{equation}
This is the UMVUE because it is a linear combination of UMVUE estimators~\cite[Corollary 4.10]{umvue}. The UMVUE for the powers of $s$ are~\cite{Glasser1962}
\begin{equation}\label{eq:umvue_sn}
    \UMVUE{s^n} =
    \begin{cases}
    \left(\frac{\nLHC}{\nMC}\right)^n \frac{k!}{(k - n)!} =
    f^n \frac{k!}{(k - n)!} & k \ge n\\
    0 & k < n\\
    \end{cases}.
\end{equation}
Performing the substitution in \cref{eq:subst} yields 
\begin{align}
    \UMVUE{\pmfPoisson{o}{s}} &= \sum_{i=0}^{k - o} \frac{(-1)^i f^{o + i} k!}{i!o! (k - o - i)!} \\
    &= \comb{k}{o} \, f^o  \, \sum_{i=0}^{k - o} \comb{k - o}{i} (-1)^i f^{i}  \\
    &= \comb{k}{o} \, f^o  \, (1 - f)^{k - o} 
    = \pmfBinom{o}{k}{f}.
\end{align}
Thus matching \cref{eq:UMVUE_b0}, the only difference being that we have $f$ whereas \refcite{Glasser1962} has $1/n$ and we may have $f > 1$ whereas $1/n \le 1$.

For the $b > 0$ case, consider the fact that we may write the exact likelihood as
\begin{equation}
 \like = \pmfPoisson{o}{s + b} = \sum_{i=0}^o \pmfPoisson{o - i}{b} \pmfPoisson{i}{s}.
\end{equation}
Thus by linearity of the UMVUE we have
\begin{equation}
 \UMVUE{\pmfPoisson{o}{s + b}} =
 \sum_{i=0}^o \pmfPoisson{o - i}{b} \UMVUE{\pmfPoisson{i}{s}} =
 \sum_{i=0}^o \pmfPoisson{o - i}{b} \pmfBinom{i}{k}{f},
\end{equation}
matching \cref{eq:UMVUE}. As previously, this is the UMVUE because it is a linear combination of UMVUE estimators.

\section{No unbiased estimator for fixed number of MC events}\label{app:no_unbiased_binomial}

Expanding in powers of the unknown efficiency, $\epsilon$, shows that no unbiased estimator can exist if we simulated a fixed number of MC events such that the number of signal events followed a binomial. The Poisson likelihood contains all powers of $\epsilon$ through the exponential term $e^{-\nLHC \epsilon}$. The expectation of any estimator (which cannot depend explicitly on $\epsilon$) from a binomial $\Binom{\nMC}{\epsilon}$ contains at most $\epsilon^{\nMC}$. That is, we would require an estimator $\hat L$ such that
\begin{equation}
   \sum_{k=0}^{\nMC} \hat L \, \pmfBinom{k}{\nMC}{\epsilon} = \pmfPoisson{o}{b + \epsilon \nLHC}
\end{equation}
but the right-hand side contains all powers of $\epsilon$, whereas the left-hand side contains at most $\epsilon^{\nMC}$. 

\thirdrevise{We can quantify the bias in an estimator that uses a fixed number of MC events as follows. Suppose we had an estimator that was an unbiased estimator of the Poisson likelihood truncated so that it contained powers no greater than $\epsilon^{\nMC}$. The remaining $\epsilon^{\nMC + 1}$  term and higher powers (for which there can be no unbiased estimator) would lead to bias
\begin{equation}
    \langle \hat L \rangle - \pmfPoisson{o}{b + \epsilon \nLHC} \approx  \fourthrevise{-} (\epsilon \nLHC)^{\nMC + 1} \, \sum_{i = 0}^o \frac{ b^{o -i} e^{-b}}{i! (\nMC +1 - i)! (o - i)!}.
\end{equation}
Thus, the bias would decay factorially with increasing $\nMC$.}

\section{Public computer codes}\label{sec:codes}

We provide a simple implementation of the UMVUE as a public library at \href{https://github.com/xhep-lab/ideal}{\texttt{github.com/xhep-lab/ideal}}.

\subsection{\code{C++} implementation}\label{sec:cpp}

We provide a simple header-only library \code{ideal.hpp} providing the function:
\begin{lstlisting}[language=C++,numbers=none]
double umvue_poisson_like(int k, double b, int o, int n_mc, double n_exp)
\end{lstlisting}
in the namespace \code{ideal}. This function returns an unbiased estimate of a Poisson likelihood when \code{k} events were simulated from an expected \code{n_mc} simulations, and \code{o} events were observed in a sample corresponding to \code{n_exp} events.

For completeness we provide functionality for drawing the number of MC simulations, 
\begin{lstlisting}[language=C++,numbers=none]
int umvue_draw_n_mc(double lambda_mc)
\end{lstlisting}
and a templated function 
\begin{lstlisting}[language=C++,numbers=none]
template<typedef engine_type>
int umvue_draw_n_mc(double lambda_mc, engine_type engine)
\end{lstlisting}
permitting greater control over random number generation. The inferred \code{engine_type} should be a random number engine, e.g., an \code{std::mt19937} from the \code{C++11} \href{https://en.cppreference.com/w/cpp/header/random}{standard library header \code{<random>}}.

The basic usage would be
\begin{colornum}
\begin{lstlisting}[language=C++]
|\color{purple}\#include| |\color{magenta}<iostream>|
|\color{purple}\#include| "ideal.hpp"

int main(int, char**) {
    std::cout << ideal::umvue_poisson_like(1000, 10, 20, 10000, 100);
    std::cout << ideal::umvue_draw_n_mc(1000);
}
\end{lstlisting}
\end{colornum}
See \code{example.cpp}. This example program can be built by entering \code{make example}.

\subsection{\code{Python} bindings}

We supply \code{Python} bindings for the \code{C++} functions \code{umvue_poisson_like} and \code{umvue_draw_n_mc} using \code{pybind11}~\cite{pybind11}. The \code{ideal} module can be built and installed by \code{make}. The equivalent example program would be:
\begin{colornum}
\begin{lstlisting}[language=Python]
import ideal

if __name__ == "__main__":
    print(ideal.umvue_poisson_like(1000, 10, 20, 10000, 100))
    print(ideal.umvue_draw_n_mc(1000))
\end{lstlisting}
\end{colornum}

\subsection{Implementation in \code{GAMBIT}}

We use code similar to that in \cref{sec:cpp} as part of the \code{ColliderBit} module of \code{GAMBIT} and implement an option to enable the UMVUE. The UMVUE may be turned on by setting the estimator in the rules governing the MC generation of collider events in the \code{yaml} input file:
\begin{lstlisting}[language=bash,keywords={Rules,capability,function,options}]
Rules:
  - capability: RunMC
    function: operateLHCLoop
    options:
      LHC_13TeV:  # or LHC_8TeV
        poisson_estimator: "UMVUE"  # by default, "MLE"
        mean_nEvents: 1000
\end{lstlisting}
We anticipate that other estimators could be added in the future. The mean number of MC events can be controlled directly by setting \code{mean_nEvents}, as in the snippet above, or by setting the ratio \code{mean_relative_nEvents}:
\begin{lstlisting}[language=bash,keywords={Rules,capability,function,options}]
Rules:
  - capability: RunMC
    function: operateLHCLoop
    options:
      LHC_13TeV:  # or LHC_8TeV
        poisson_estimator: "UMVUE"  # by default, "MLE"
        mean_relative_nEvents: 2
\end{lstlisting}
If both \code{mean_nEvents} and \code{mean_relative_nEvents} are specified, \code{GAMBIT} throws a user input error. The default behavior is to run with $\code{mean_relative_nEvents} = 1$. The MLE estimator can also be used with the new \code{mean_nEvents} and \code{mean_relative_nEvents} settings, as well as with the existing \code{min_nEvents} and \code{max_nEvents} settings.

Our implementation of the \code{mean_relative_nEvents} option involved two complications. First, we wish to use the same set of MC events to compute selection efficiencies for multiple searches and combine likelihood estimators through \cref{eq:product}. Whilst these different searches would share an identical cross-section, they could be associated with different integrated luminosities. We thus set the expected number of MC events through,
\begin{equation}\label{eq:r_mc}
    \nMC = \code{mean_relative_nEvents} \times \max \nLHC = \code{mean_relative_nEvents} \times \sigma \times \max \lumi,
\end{equation}
where we maximize across searches sharing the same MC events.

Using \cref{eq:r_mc} to set the expected number of MC events requires computation of the cross-section before event generation. Thus, we separate cross-section computation and event generation in \code{GAMBIT}. Initial cross-section estimates are performed by simulating the hard scatter in \code{Pythia} without showering or hadronisation. Separate events are generated with showering and hadronisation during the \code{ColliderBit} main event loop for calculating signal estimates in collider analyses. In the near future, we will also support extracting these initial estimates from external cross-section calculators.

The UMVUE estimator interacts with other \code{ColliderBit} settings as follows. The number of MC events generated by \code{ColliderBit} was previously controlled by setting a minimum and a maximum number of MC events through \code{min_nEvents} and \code{max_nEvents}, respectively. These settings allow \code{ColliderBit} to end event generation when \code{max_nEvents} events have been generated or when anywhere between \code{min_nEvents} and \code{max_nEvents} have been generated and convergence criteria are met. These options are ignored when enabling the UMVUE estimator. By clipping the Poisson distributed number of MC events to lie between \code{min_nEvents} and \code{max_nEvents}, they would introduce bias into the estimator. In the event that the UMVUE estimator is used with these settings, \code{ColliderBit} will throw an error.

Second, in \code{ColliderBit}, the uncertainty in a predicted signal plus background $(s+b)$ can be described by a normal or log-normal probability distribution with variance $\gamma^2$ for each signal region, set by the combination of experiment-provided background uncertainty and MC signal uncertainty. In \code{ColliderBit}, these nuisances $(\gamma)$ can be either profiled or marginalized. However, since the UMVUE was constructed for Bayesian computation, only marginalization is implemented for the UMVUE estimator. An appropriate error is thrown when attempting to profile these uncertainties and use the UMVUE estimator.

The marginalization of these nuisances in \code{ColliderBit} is performed by Monte Carlo integration. The resulting estimate is thus unbiased. As evaluating the integrand is computationally cheap compared to collider simulation, the noise can in any case be made negligible.

\end{document}